# Measuring publication relatedness using controlled vocabularies

Emil Dolmer Alnor

*ea@ps.au.dk*
0000-0002-8536-9442
Danish Centre for Studies in Research and Research Policy, Aarhus University, Denmark

ABSTRACT: Measuring the relatedness between scientific publications has important applications in many areas of bibliometrics and science policy. Controlled vocabularies provide a promising basis for measuring relatedness because they address issues that arise when using citation or textual similarity to measure relatedness. While several controlled-vocabulary-based relatedness measures have been developed, there exists no comprehensive and direct test of their accuracy and suitability for different types of research questions. This paper reviews existing measures, develops a new measure, and benchmarks the measures using TREC Genomics data as a ground truth of topics. The benchmark test show that the new measure and the measure proposed by Ahlgren et al. (2020) have differing strengths and weaknesses. These results inform a discussion of which method to choose when studying interdisciplinarity, information retrieval, clustering of science, and researcher topic switching.

## 1. Introduction

Measuring the relatedness between scientific publications has important applications in many areas of bibliometrics and science policy. It is central in measuring interdisciplinarity (Rafols & Meyer, 2010; Stirling, 2007), aiding information retrieval (Lin & Wilbur, 2007), clustering science into topics (Waltman & Van Eck, 2012), and analysing the causes and consequences of researchers adapting their research trajectories (Hill et al., 2021; Myers, 2020). The relatedness between scientific publications is often measured using their citation relation, textual similarity, or a combination of the two (Waltman et al., 2020). While applicable to many publications, these approaches have some known issues, e.g., citation time-lags, synonyms, homonyms, and polysemy (Glänzel & Thijs, 2017, pp. 1072-1073).

Using controlled vocabularies to measure relatedness addresses these issues, because publications are often indexed with keywords shortly after publication and because controlled vocabularies provide consistency and uniqueness in how refer to concepts. Medical Subject Headings (MeSH) is a controlled vocabulary used to index publications in PubMed, and several measures that use MeSH to measure publication relatedness have been developed (Ahlgren et al., 2020; Boudreau et al., 2016; Zhou et al., 2015).

This paper provides a critical review of these measures, argues that they lack precision, and therefore develops a new measure. The paper then uses data from TREC 2006 Genomics Track to test the measures' accuracy, and argues that this test provides more insight into their accuracy than previous tests (e.g.,Waltman et al., 2020). The benchmark test show that the new measure and the measure proposed by Ahlgren et al. (2020) are the best performing, but that they have differing strengths and weaknesses. These results inform a discussion of which measure is most useful when studying interdisciplinarity, information retrieval, clustering of science, and researcher topic switching. The results and discussions provided by this paper is therefore relevant to researchers studying these types of questions.

## 2. Measuring publication relatedness using controlled vocabularies

The existing measures observe either co-occurrence or distance between the terms which publications are indexed with. All reviewed measures use MeSH as their controlled vocabulary, however, the methods could be applied to any controlled vocabulary, thesaurus, or ontology as long as it describes the relationship between its terms.

MeSH contains more than 30.000 terms which describe scientific concepts, and which are hierarchically organized (U.S. National Library of Medicine, 2024). For example, 'Heart Arrest' is a subterm to 'Heart Diseases'. Publications in PubMed are typically indexed with 15 MeSH-terms, with typically 4 terms designated as 'major', meaning that they reflect the primary topics of the publication. The remaining 'minor' terms reflect concepts of less importance to the publication. Around 90% of terms have qualifiers, describing the specific aspect of the concept. For example, 'rehabilitation' can be used as a qualifier for 'Heart Arrest'. Until 2022 MeSH indexing was done by human field experts and since April 2022 human field experts have curated and reviewed machine-based indexing.

*2.1. Co-occurrence based approaches*

The simplest co-occurrence approach is to measure the cosine similarity between vectors of MeSH-terms, which resembles the approach applied by Boudreau et al. (2016). The intuition is that the more MeSH-terms two publications have in common relative to their total number of MeSH-terms, the more similar they are. Formally, let publication $p_a$ be indexed with a set of MeSH-terms, $P_a$. Represent $p_a$ with vector $\mathbf{v}_a = [v_{a1}, v_{a2}, \ldots v_{an}]$, where $n$ is the number of unique MeSH-terms, and $v_{ai} = 1$ if $p_a$ is indexed with MeSH-term $i$, i.e., $m_i \in P_a$, and 0 otherwise. Then the similarity between publications $p_a$ and $p_b$ is defined as the cosine similarity between their vectors, $\mathbf{v}_a$ and $\mathbf{v}_b$:

$$sim_{Boudreau}(p_a, p_b) = \frac{\mathbf{v}_a \cdot \mathbf{v}_b}{\|\mathbf{v}_a\| \|\mathbf{v}_b\|} \tag{1}$$

While this should capture some of the relatedness between $p_a$ and $p_b$, there are some inaccuracies: Two publications having a very frequent MeSH-term in common is deemed as similar as two publications having a very rare MeSH-term in common, and the method does not utilize the information in the distinction between major and minor MeSH-terms nor their qualifiers.

The co-occurrence-based approach suggested by Ahlgren et al. (2020) addresses all these problems. The intuition is that the more information two publications share relative to their total information, the more similar they are. Ahlgren et al. (2020) start by defining a weight (information content, IC) for each MeSH-term, which is low if the frequency of the MeSH-term and its descendants is high. Formally, let $\#(m_i)$ denote the frequency of MeSH-term $i$ in the data set, let $D_i$ denote the set of MeSH-terms that are descendants to $m_i$ either direct or indirect, and let $n$ denote the number of unique MeSH-terms. Then the IC of $m_i$ is defined as

$$\text{IC}(m_i) = -\log\left(\frac{\#(m_i) + \sum_{d \in D_i} \#(d)}{\sum_{k=1}^{n} \#(m_k) + \sum_{d \in D_k} \#(d)}\right) \tag{2}$$

Now, with $q$ denoting the number of unique qualifiers and $m_{i_q}$ denoting MeSH-term $i$ with qualifier $q$, represent publication $p_a$ by a vector $\mathbf{v}_a = [v_{a1}, v_{a1_1}, v_{a1_2}, \ldots v_{a1_q}, v_{a2}, v_{a2_1}, v_{a2_2}, \ldots v_{a2_q}, \ldots v_{am_q}]$. To utilize the qualifiers, let $v_{ai_q} = 1$ if $m_{i_q} \in P_a$ and 0 otherwise. To utilize the IC of MeSH-terms and the distinction between major and minor terms let

$$v_{ai} = \begin{cases} 0 \text{ if } m_i \notin P_a \\ IC(m_i) \text{ if } m_i \text{ is a minor term in } P_a \\ IC(m_i) \times 2 \text{ if } m_i \text{ is a major term in } P_a \end{cases} \quad (3)$$

Again, the similarity between publications is then defined as the cosine similarity between their vectors.

While this addresses the inaccuracies in $sim_{Boudreau}$, a substantial problem remains. Consider three publications, $p_a$, $p_b$, and $p_c$, which all share the two MeSH-terms 'Hospitals' and 'Health Personnel' but differ in the third term they have: $p_a$ has 'Heart Arrest', $p_b$ has 'Heart Failure', while $p_c$ has 'Horse diseases'. Intuitively, $sim(p_a, p_b) > sim(p_a, p_c)$, because 'Heart Arrest' is more related to 'Heart Failure' than to 'Horse Diseases'. The co-occurrence-based approaches, however, only observe whether publications share the exact same MeSH-terms. Therefore $sim_{Boudreau}(p_a, p_b) = sim_{Boudreau}(p_a, p_c) = sim_{Boudreau}(p_b, p_c)$. In $sim_{Ahlgren}$ it could be worse. Since the dot products in all three publication pairs' cosine similarity are equal, their similarity depends on the product of their vector magnitudes which is decided by the IC of the MeSH-terms. Specifically, a large IC leads to a large denominator, and therefore, $IC(m_{Heart\ Failure}) > IC(m_{Horse\ diseases}) \Rightarrow sim_{Ahlgren}(p_a, p_b) < sim_{Ahlgren}(p_a, p_c)$.

*2.2. Distance-based approaches*
A solution to this problem is to measure relatedness by the distance between MeSH-terms. This approach involves first a method for measuring MeSH-term distance and secondly a method for aggregating MeSH-term distance to publication distance.

*2.2.1 Measuring term distance*
This paper suggests two methods for defining MeSH-term distance. Both require introducing an additional detail about the MeSH-terms: Each MeSH-term has one or more tree numbers, which show its position(s) in the MeSH-tree. This reflects the fact that MeSH-terms may belong to several categories. For example, 'Rehabilitation' has several tree-numbers because it is both a therapeutic technique, an occupational speciality, and a type of health care service.

To measure MeSH-term distance, let $G_1 = (V, E)$ be an unweighted undirected graph, where $V = [m_1, m_2, \ldots, m_n]$, i.e., each unique MeSH-term constitute a vertex, and $E = [e_1, e_2, \ldots e_k]$, where $k$ is the number of parent-child relationships between the MeSH-terms. Specifically, let $e = (m_i, m_j)$ if $m_i$ is the child or parent of $m_j$. Note that MeSH-terms with several tree numbers can have several parents. The distance between two MeSH-terms is then defined as the shortest path between them.

Since the graph is unweighted, the distance between all parent-child pairs of MeSH-terms is equal. If a MeSH-term has a single child with almost equal IC as itself, it could be assumed that the distance between this parent-child pair is shorter than the distance between a parent-child pair having a large difference in information content. To test this assumption, a second method of defining MeSH-term distance is included. Let $G_{\Delta IC} = (V, E, sl)$ be a weighted undirect graph identical to $G_1$ except each edge has a weight, $sl$ (step length). Specifically, let

$$sl\big(e(m_i, m_j)\big) = |IC(m_i) - IC(m_j)| \quad (4)$$

In words, the distance between two directly related MeSH-terms is their difference in information content, and, as in $G_1$, the distance between any two MeSH-terms is the shortest path between them.[1]

*2.2.2 Aggregating term distance to publication distance*
The paper tests four methods of aggregating MeSH-term distance to determine publication distance. In the simplest approach, also applied by Zhu et al. (2009), the distance between the $p_a$ and $p_b$, is defined as:

$$dist_1(p_a, p_b) = \frac{\sum_{m_i \in P_a} \min_{m_j \in P_b} dist(m_i, m_j) + \sum_{m_j \in P_b} \min_{m_i \in P_a} dist(m_j, m_i)}{|P_a| + |P_b|} \quad (5)$$

i.e., it is the sum of minimum distances from terms in $P_a$ to terms in $P_b$ and vice versa, divided by the number of MeSH-terms in $P_a$ and $P_b$.

In this approach, minor and major MeSH-terms contribute equally to computing publication distance, which is not desirable. To address this problem Zhu et al. (2009) propose to exclude minor terms from $P_a$ and $P_b$ (denoted $dist_0$ here). However, this approach throws away the information inherent in the minor terms. Instead, this paper tests a different method which applies a weight, $w$, to major terms:

$$dist_w(p_a, p_b) = \frac{\sum_{m_i \in P_a} \min_{m_j \in P_b} dist(m_i, m_j) \times w_{m_i} + \sum_{m_j \in P_b} \min_{m_i \in P_a} dist(m_j, m_i) \times w_{m_j}}{\sum_{m_i \in P_a} w_{m_i} + \sum_{m_j \in P_b} w_{m_j}} \quad (6)$$

i.e., it is the sum of weighted minimum distances divided by the sum of weights. Minor terms receive weight 1 and the paper test two weighting factors: $dist_2$ and $dist_3$ where major terms receive weight 2 and 3 respectively, e.g., in $dist_2$ let $w_{m_i} = 2$ if $p_a$ has $m_i$ as a major term.

Summing up, eight distance measures are tested, which are subscripted by the weights applied and superscripted by the method used to calculate path length, e.g., $dist_3^1$ means that major MeSH-terms receive weight 3, and each step has length 1.

## 3. Evaluating relatedness measures
This section explains the tests used to evaluate the relatedness measures. These differ from the typical test, which uses the relatedness measure in a clustering algorithm and then compares the retrieved topics to some golden standard of topics (Waltman et al., 2020; Zhu et al., 2009). The reason for this deviation is that the typical test only indirectly evaluates the relatedness measure itself. Clustering solutions are produced by both the relatedness measure used in the similarity matrix and the algorithm used to cluster the similarity matrix (Held, 2022; Šubelj et al., 2016). Therefore, the accuracy of a clustering solution cannot be attributed solely to the relatedness measure. While using different relatedness measures in a clustering algorithm and assessing the accuracy of the resulting clusters provides insight into the accuracy of those relatedness measures within that specific clustering algorithm, it provides less insight into their accuracy within other clustering algorithms and even less insight into the accuracy of the relatedness measures themselves. As highlighted in the introduction, relatedness measures have many applications beyond clustering.

---

[1] Note that these two measures of MeSH-term distance resemble Zhu et al. (2008)'s equation 1 and 3, respectively. The difference is that Zhu et al. (2008) conceptually start from a tree and look for the closest common ancestor, whereas this paper starts from a network and looks for the shortest path. MeSH-terms typically have more than 1 node representation (MeSH 2024 tree has median = 2, mean = 2,1), and the number of branch merges is therefore so high, that a network view of the MeSH-hierarchy may be assumed to be more appropriate. Testing this assumption is work in progress, which will be presented at the conference if the paper is accepted.

*3.1. TREC Genomics*

The benchmark data comes from the TREC Genomics Track 2006 (Hersh et al., 2006). Here, teams of information scientists were given 28 information needs (referred to as topics below) and the task to nominate relevant text passages from a set of approximately 170,000 publications. All topics concerned the role of genes in biological processes or diseases, such as 'What is the role of PrnP in mad cow disease?'. Each of the 92 teams submitted 1000 text passages spread across the 28 topics. Subsequently, field experts evaluated these passages as either '0 – not relevant', '1 – possibly relevant', or '2 – relevant' to the topic. To get relevance judgements on publication level, the relevance of a publication to a topic is here defined as the maximum relevance of any of its passages to that topic. The tests below only include topics where at least 10% of the nominated publications have relevance 1 or 2. Table 1 shows the distribution of relevance judgements across these topics.

Table 1: Distribution of relevance judgements across topics.

| Topic | 0 - Not relevant | 1 – possibly relevant | 2 - relevant | Ratio of 0's |
|---|---|---|---|---|
| 181 | 122 | 38 | 182 | 0.357 |
| 160 | 125 | 60 | 138 | 0.387 |
| 186 | 230 | 29 | 163 | 0.545 |
| 172 | 356 | 27 | 207 | 0.603 |
| 163 | 269 | 94 | 72 | 0.618 |
| 182 | 411 | 31 | 44 | 0.846 |
| 167 | 360 | 9 | 47 | 0.865 |
| 168 | 280 | 21 | 22 | 0.867 |
| 169 | 422 | 20 | 44 | 0.868 |

*3.2. Benchmark test 1*

The first test evaluates the relatedness measures' ability to compute higher relatedness to pairs of publications both deemed '2 – relevant' to the same topic (rr's), compared to the relatedness scores between pairs where one publication is '0 - not relevant' and the other is '2 - relevant' (nr-r's). Using a medical analogy, this test resembles measuring differences in outcomes (the relatedness score) across a control (nr-r's) and a treatment group (rr's). Since the relatedness measures produce metrics at different scales it is not valid to compare raw differences in relatedness. An often-applied solution is to calculate Cohens D. However, this metric assumes normality of distributions, and Figure 2 shows that this assumption is violated for the co-occurrence-based measures. Instead, the non-parametric Cliff (1993) delta is calculated, which is defined as:

$$d = \frac{\sum_{i,j}[x_i > x_j] - [x_i < x_j]}{mn} \quad (7)$$

where the number of rr's and nr-r's are $m$ and $n$ with similarity scores $x_i$ and $x_j$, respectively, and the Iverson brackets [] are 1 when their contents are true and 0 otherwise. The intuition behind the measure is that it quantifies the proportion of rr's with higher relatedness than nr-r's.

*3.3. Benchmark test 2*
The second benchmark test evaluates the relatedness measures' ability to correctly classify publications as belonging to a topic or not. The test assumes that a topic can be represented by a sample of publications that are judged relevant to that topic. If a relatedness measure is accurate, non-sampled publications also judged relevant should have a high relatedness score to the sampled publications.

Formally, the test does 30 iterations of the following for each of the topics. First, to define the sets $R_t$ and $NR_t$, which does and does not represent topic *t*, respectively, it samples 10 publications which are relevant and 10 publications which are not relevant to topic *t*. Let $p_i \notin (R_t \cup NR_t)$ be an unsampled publication which has received a judgement as either relevant or not relevant to topic *t*. Let $\max(rel(p_i, R_t))$ denote the maximum relatedness of $p_i$ to any $p_r \in R_t$ and let $\max(rel(p_i, NR_t))$ denote the maximum relatedness of $p_i$ to any $p_{nr} \in NR_t$. Then, the relevance judgement of $p_i$ to topic *t* according to relatedness measure *r* is defined as:

$$rj_{itr} = \begin{cases} 2 \text{ if } \max(rel_r(p_i, R_t)) > \max(rel_r(p_i, NR_t)) \\ 0 \text{ if } \max(rel_r(p_i, R_t)) \leq \max(rel_r(p_i, NR_t)) \end{cases} \quad (8)$$

That is, if relatedness measure *r* shows that the publication most related to $p_i$ is in the sample of publications relevant to topic *t*, then $p_i$ is judged relevant to topic *t* according to relatedness measure *r*. The relatedness measures' relevance judgements are then compared with the experts' relevance judgements to calculate the number of true positives (TP), false positives (FP), true negatives (TN), and false negatives (FN). These are used to calculate precision, recall, and Matthews (1975) correlation coefficient ($\varphi$), the latter defined as:

$$\varphi = \frac{TP \times TN - FP \times FN}{\sqrt{(TP + FP)(TP + FN)(TN + FP)(TN + FN)}} \quad (9)$$

Among other reasons, $\varphi$ is a good metric for evaluating binary classification systems because it includes all categories of the confusion matrix, which, e.g., the $F_1$-score does not (Chicco & Jurman, 2023).

## 4. Results
Figure 1 shows the density distribution of the relatedness measures. The distance measures follow a normal-like distribution, whereas the similarity measures follow a right-skewed distribution in which most publication pairs receive either 0 or very low relatedness. The latter is surprising since all publications in both the nr-r's and rr's were nominated by information scientists to be relevant to the same topic. This indicates that the co-occurrence-based measures lack discriminatory power in distinguishing the relatedness of pairs of only remotely related publications.

Table 2 shows that within the distance measures, $dist_3^{\Delta IC}$ receives the highest performance metrics on both test 1 and test 2. As expected, $sim_{Ahlgren}$ performs better than $sim_{Boudreau}$. On test 1, $sim_{Ahlgren}$ performs best with a Cliff's delta of 0.41, which is 0.09 more than that of $dist_3^{\Delta IC}$. Turning to test 2, $sim_{Ahlgren}$ has a precision of 32.3%, which is 1 percentage point more than

Figure 1: Density plots of relatedness measures.

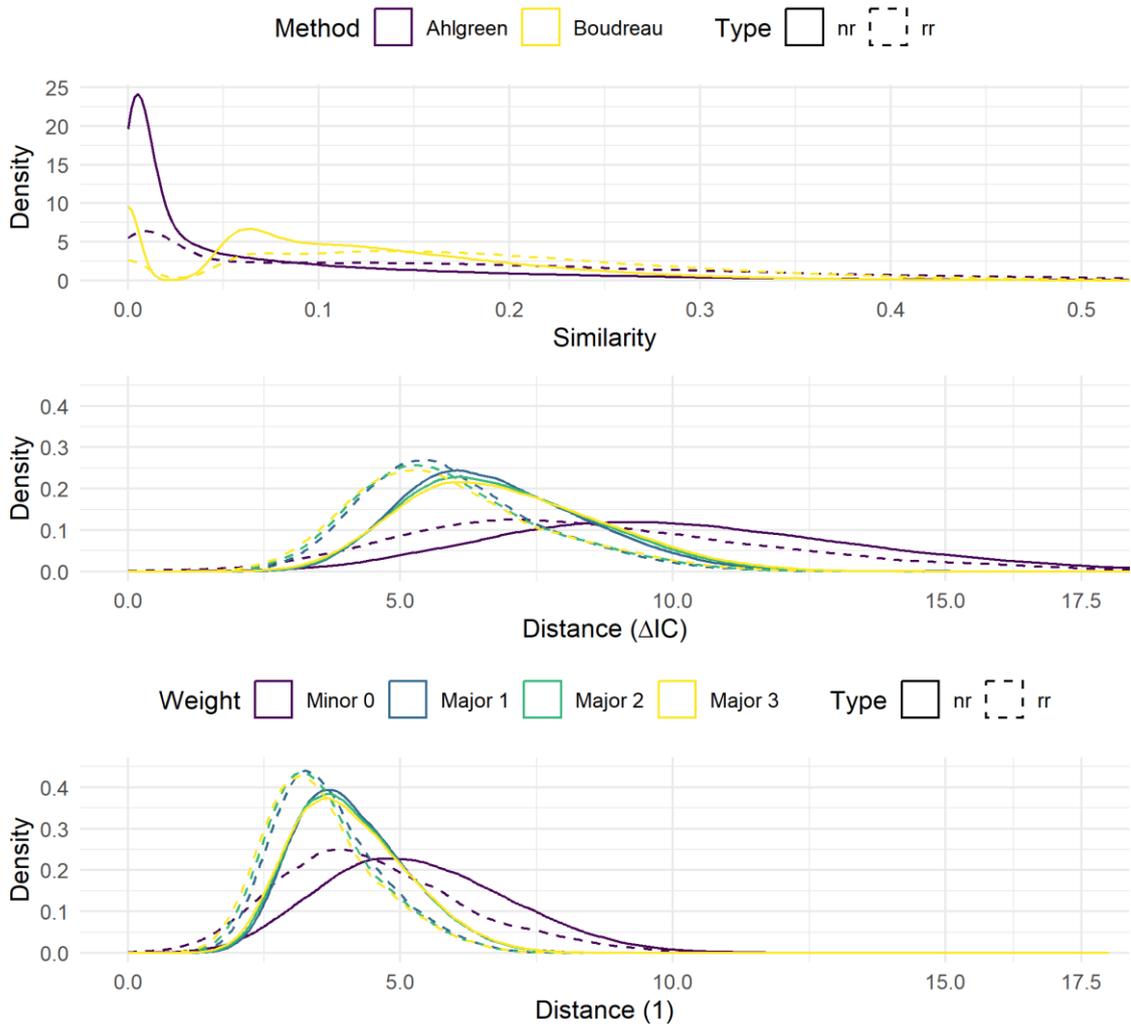

Note: The x-axes are truncated. Approximately 1% of pairs have similarities (distances) larger than 0.5 (17.5).

Table 2: Benchmark tests of relatedness measures.

| | Mean distance/similarity | | | Positives | | Negatives | | | | |
|---|---|---|---|---|---|---|---|---|---|---|
| Method | nr-r | r-r | Cliff's d | True | False | True | False | Precision | Recall | $\varphi$ |
| $dist_1^{\Delta IC}$ | 6.791 | 5.994 | 0.276 | 17,870 | 40,431 | 33,729 | 7,000 | 30.7% | 71.9% | 0.153 |
| $dist_2^{\Delta IC}$ | 6.892 | 5.956 | 0.306 | 18,332 | 40,409 | 33,751 | 6,538 | 31.2% | 73.7% | 0.170 |
| $dist_3^{\Delta IC}$ | 6.954 | 5.928 | 0.320 | 18,460 | 40,432 | 33,728 | 6,410 | 31.3% | 74.2% | 0.174 |
| $dist_0^{\Delta IC}$ | 10.225 | 8.615 | 0.275 | 17,817 | 41,328 | 32,832 | 7,053 | 30.1% | 71.6% | 0.141 |
| $dist_1^1$ | 4.114 | 3.694 | 0.235 | 17,217 | 39,058 | 35,102 | 7,653 | 30.6% | 69.2% | 0.145 |
| $dist_2^1$ | 4.093 | 3.606 | 0.268 | 17,739 | 39,574 | 34,586 | 7,131 | 31% | 71.3% | 0.158 |
| $dist_3^1$ | 4.079 | 3.548 | 0.286 | 18,042 | 39,844 | 34,316 | 6,828 | 31.2% | 72.5% | 0.166 |
| $dist_0^1$ | 5.306 | 4.487 | 0.276 | 17,529 | 41,555 | 32,605 | 7,341 | 29.7% | 70.5% | 0.128 |
| $sim_{Ahlgren}$ | 0.067 | 0.165 | 0.407 | 19,216 | 40,297 | 33,863 | 5,654 | 32.3% | 77.3% | 0.203 |
| $sim_{Boudreau}$ | 0.118 | 0.178 | 0.328 | 18,050 | 39,696 | 34,464 | 6,820 | 31.3% | 72.6% | 0.168 |

Note: n: column 2 = 228,691; column 3 = 66,208; column 4 = 294,899.

$dist_3^{\Delta IC}$, and a recall of 77.3%, which is 3.1 percentage points higher than $dist_3^{\Delta IC}$, which has a recall of 74.2%. $sim_{Ahlgren}$ has $\varphi = 0.203$ which is 0.029 higher than that $dist_3^{\Delta IC}$. This indicates that $sim_{Ahlgren}$ has the overall best performance on test 2.

## 5. Discussion and future work

The differential performance of $dist_3^{\Delta IC}$ and $sim_{Ahlgren}$, along with their density distributions indicates their suitability for different research questions. While the tests indicated that $sim_{Ahlgren}$ is better at discerning the relatedness between highly related publication pairs, $dist_3^{\Delta IC}$ may be superior at discerning the relatedness between distantly related publications due to its normal-like distribution, contrasting with $sim_{Ahlgren}$ right-skewed distribution.

Therefore, when studying researchers adapting research trajectories (Hill et al., 2021; Myers, 2020) or interdisciplinarity (Rafols & Meyer, 2010; Stirling, 2007), method selection may depend on the degree of adaptation or the granularity of interdisciplinarity being studied. $dist_3^{\Delta IC}$ may be better when studying substantial career changes or interdisciplinarity across disciplines, while $sim_{Ahlgren}$ may be better when studying incremental career changes or interdisciplinarity across specialities. In large-scale clustering, computational demands necessitate sparsening the publication similarity matrices anyway (Boyack et al., 2011; Waltman et al., 2020), i.e., only highly related publication pairs are considered, and therefore $sim_{Ahlgren}$ may be chosen for its accuracy. In information retrieval $sim_{Ahlgren}$ is the superior method. In econometric studies that use parametric tests (e.g., Myers, 2020), $dist_3^{\Delta IC}$ is advantageous because it requires no transformation to satisfy normality assumptions.

The generalisability of this work is limited by only using TREC Genomics 2006 as benchmark data and by only using MeSH as the controlled vocabulary. Future research could expand the tests to include data from TREC Genomic 2005 and 2004, test other controlled vocabularies, and compare them against citation-based, text-based, or hybrid relatedness measures.

**Open science practices**
The data used in the paper is publicly available from The National Center for Biotechnology Information (NCBI). The paper uses the open-source statistical software R, and the script is available at https://github.com/EDAlnor/MeSH-relatedness. All results can be replicated by running the script, which will download, process, summarise and visualise the data.

**Acknowledgments**
The author thanks Jens Peter Andersen and other colleagues at the Danish Centre for Studies in Research and Research Policy for useful comments and suggestions. The usual disclaimer applies.

**Competing interests**
The author has no competing interests.

**Funding information**
No funding was received for this study.